\documentclass[aps,prl,twocolumn,preprintnumbers,10pt,showpacs]{revtex4-1}

\usepackage{amsmath,bm,amssymb}

\usepackage{graphicx}
\usepackage{psfrag}
\usepackage[caption=false]{subfig}
\usepackage{color,xcolor}
\definecolor{urlblue}{rgb}{0.2,0.4,0.7}
\definecolor{citegreen}{rgb}{0,0.6,0.2}
\definecolor{linkred}{rgb}{0.9,0.2,0.1}
\usepackage{hyperref}
\hypersetup{
colorlinks=true, citecolor=citegreen, linkcolor=blue,urlcolor=urlblue}

\usepackage{slashed}
\usepackage{array}
\newcolumntype{P}[1]{>{\centering\arraybackslash}p{#1}}

\allowdisplaybreaks

\def\zo{\overline{z}_1}
\def\zt{\overline{z}_2}

\def\Aob{\overline A_1^I}
\def\Atb{\overline A_2^I}
\def\Athb{\overline A_3^I}

\def\Ao{A_1^I}
\def\At{A_2^I}

\def\Bo{B_1^I}
\def\Bt{B_2^I}

\def\Dobd{\overline D_{d,1}^I}

\def\Dtbd{\overline D_{d,2}^I}

\def\btob{\overline \beta_1}
\def\bttb{\overline \beta_2}

\def\lfr{L_{fr}}
\def\lw{\ln(1-\omega)}
\def\w{\omega}
\def\lqr{L_{qr}}

\begin{document}

\preprint{IMSc/2017/08/07}

\title{Threshold resummation of the rapidity distribution for Higgs production at NNLO+NNLL
}

\author{Pulak Banerjee$^{a}$}\email{bpulak@imsc.res.in}
\author{Goutam Das$^{a}$}\email{goutam@imsc.res.in}
\author{Prasanna K. Dhani$^{a}$}\email{prasannakd@imsc.res.in}
\author{V. Ravindran$^{a}$}\email{ravindra@imsc.res.in}

\affiliation{$^a$ The Institute of Mathematical Sciences, HBNI, Taramani,
 Chennai 600113, India} 

%
%
%

\date{\today}

\begin{abstract}
We present a formalism that resums threshold-enhanced logarithms to all orders in perturbative QCD for the rapidity distribution of any colorless particle produced in hadron colliders. We achieve this by exploiting the factorization properties and K+G equations satisfied by the soft and virtual parts of the cross section. We compute for the first time compact and most general expressions in two-dimensional Mellin space for the resummed coefficients.  Using various state-of-the-art multiloop and multileg results, we demonstrate the numerical impact of our resummed results up to next-to-next-to-leading order for the rapidity distribution of the Higgs boson at the LHC. We find that inclusion of these threshold logs through resummation improves the reliability of 
perturbative predictions.
\\

\hspace{2.cm}{\centering {\it \large This article is dedicated to the memory of Jack Smith.}}
\end{abstract}


\maketitle
\textit {Introduction}.---With the successful running of the LHC at CERN and precise theoretical predictions
from various state-of-the-art computations,    
we can now test the Standard Model (SM) of particle physics with unprecedented
accuracy and also severely constrain many physics beyond the SM (BSM) scenarios.  The spectacular discovery 
\cite{Aad:2012tfa,*Chatrchyan:2012xdj} of a scalar
particle and the most precise prediction on its production cross section \cite{Anastasiou:2015ema}  
improved our understanding of the symmetry-breaking mechanism, namely, the Higgs mechanism.  
%
%
The copious production of vector bosons $Z$s and $W^{\pm}$s and lepton pairs at the LHC through 
Drell-Yan (DY) process \cite{Drell:1970wh}, 
which are used to precisely measure the parton distribution functions (PDFs) \cite{Gao:2013xoa,*Harland-Lang:2014zoa,*Ball:2014uwa,*Butterworth:2015oua,*Alekhin:2017kpj} are also very important to study.

While inclusive rates are important for any phenomenological study, 
the differential cross sections often carry more information on the nature of interaction and  
quantum number of particles produced in the hard collisions. 
Rapidity distributions of Drell-Yan pair \cite{CMS:2014jea}, $Z$ boson \cite{Affolder:2000rx}, and charge 
asymmetries of leptons in $W^{\pm}$ boson decays \cite{Abe:1998rv} 
are already used to measure PDFs. 
Possible excess events in these distributions can hint at BSM physics, 
namely, R-parity violating supersymmetric models  \cite{Affolder:2001ha},
models with $Z'$ or with contact interactions, and large
extra-dimension models  \cite{ArkaniHamed:1998rs,*Randall:1999ee}.
Like in DY, measurements of transverse momentum and rapidity distributions of the Higgs boson will be 
very useful to study the properties of the Higgs boson and its couplings.
Theoretical predictions for inclusive production  
\cite{Dawson:1990zj,*Djouadi:1991tka,*Spira:1995rr,*Harlander:2001is,*Catani:2001ic,*Harlander:2002wh,*Anastasiou:2002yz,*Ravindran:2003um,*KubarAndre:1978uy,*Altarelli:1978id,*Humpert:1980uv,*Matsuura:1987wt,*Matsuura:1988sm,*Hamberg:1990np} as well as the rapidity distribution \cite{Anastasiou:2003yy,*Anastasiou:2004xq}
of dileptons in DY production 
and the Higgs bosons in gluon-gluon fusion have been known to next-to-next-to-leading order
(NNLO) in perturbative QCD for long time. A few years back, a complete next-to-next-to-next-to-leading-order (N$^3$LO)
prediction \cite{Anastasiou:2015ema} for inclusive Higgs production became available after its 
soft-plus-virtual (SV) contributions (N$^3$LO$_{SV}$) 
were computed in Ref.~\cite{Anastasiou:2014vaa}, see also Refs.~\cite{Moch:2005ky,*Laenen:2005uz,*Idilbi:2005ni,Ravindran:2005vv,Ravindran:2006cg} for earlier works and Ref.~\cite{Kumar:2014uwa, *Ahmed:2014cha} for Higgs
productions in other channels at N$^3$LO$_{SV}$ and Ref.~\cite{Ahmed:2015sna} for a renormalization group improved prediction to all orders for $g g \rightarrow H$.  For DY, so far, only  N$^3$LO in the SV approximation is known \cite{Ahmed:2014cla}, see also Ref.~\cite{Li:2014bfa,*Catani:2014uta}.    

Both inclusive and differential cross sections are often plagued with large logarithms resulting from certain boundaries
of the phase space, spoiling the reliability of the fixed-order predictions.  
In the inclusive case, this happens when partonic scaling variable $z = Q^2/\hat s \rightarrow 1$, i.e., 
threshold limit, resulting from the emission of soft gluons in the DY process ($Q^2=m_{l^+l^-}^2$) and in 
Higgs production ($Q^2=M_H^2$), where  $m_{l^+l^-}$, $M_H$, and $\hat s$ are the
invariant mass of the dileptons, the mass of the Higgs boson,
and centre-of-mass energy squared of the partonic subprocess, respectively.  
One finds a similar problem when the transverse momentum of the final state becomes small. 
The resolution to this is to resum these large logs to all orders in perturbation theory.
To achieve this, several approaches exist in the literature for both inclusive rates (see Refs.~\cite{Sterman:1986aj,Catani:1989ne,*Catani:1990rp} for the earliest approach) as well as for transverse momentum distributions 
\cite{Bozzi:2007pn,*Bozzi:2008bb,*Bozzi:2010xn,*Catani:2010pd, *Catani:2013vaa, *Monni:2016ktx, *Ebert:2016gcn, *Grazzini:2015wpa, *Ferrera:2016prr}.
Catani and Trentadue, in their seminal work \cite{Catani:1989ne}, 
demonstrated the resummation of leading large logs for the inclusive rates in 
Mellin space and extended their approach to a differential $x_F$ distribution using double Mellin moments.   
In the recent past, there have been several 
 approaches to performing threshold resummation for rapidity
distribution.  In Ref.~\cite{Laenen:1992ey,*Sterman:2000pt,*Mukherjee:2006uu,*Bolzoni:2006ky}, an appropriate Fourier transformation for the rapidity variable resums 
certain logs for the rapidity distribution, 
and in Ref.~\cite{Becher:2006nr,*Becher:2007ty,*Bonvini:2014qga}, the authors have used soft-collinear effective theory (SCET) to identify the
potential large logs that can be resummed (see also Ref.~\cite{Ebert:2017uel} for resumming
timelike logarithms using SCET). 

In Refs.~\cite{Ravindran:2005vv,Ravindran:2006cg}, one of the authors of the present article 
developed $z$-space formalism
to obtain a soft distribution function that captures 
the threshold-enhanced part of the inclusive production of any colorless particle,
using factorization properties of cross sections  
and K+G equations that the form factor as well as soft distribution function satisfy.
In Ref.~\cite{Ravindran:2006cg}, it was shown that the $N$th Mellin moment of the 
finite part of the universal soft distribution function was nothing but the threshold exponent \'{a}~la Sterman~\cite{Sterman:1986aj} and 
Catani and Trentadue \cite{Catani:1989ne,Catani:1990rp}. The same approach was 
later extended to obtain rapidity distributions of lepton pairs, Higgs boson 
\cite{Ravindran:2006bu}, and $Z$ and $W^{\pm}$ \cite{Ravindran:2007sv} using two scaling 
variables $z_1$ and $z_2$ in the threshold limit up to N$^3$LO level \cite{Ahmed:2014uya,*Ahmed:2014era}.

In this article, we derive an all-order resummed result in two-dimensional Mellin space for rapidity distribution of a colorless final state $F$ that can be produced in hadron colliders and present the numerical impact only for the production of the scalar Higgs boson at the LHC.    
We work with double Mellin variables $N_1$ and $N_2$ corresponding to $z_1$ and $z_2$ in $z$ space and demonstrate the resummation of large logarithms proportional to $\ln(N_i)$ (in z space, these correspond to plus distributions in both the variables $z_1$ and $z_2$) in the limit $N_i \to \infty$ ($z_i \to 1$). 
Our approach, while it follows Ref.~\cite{Catani:1989ne}, 
differs from Refs.~\cite{Sterman:2000pt,Bolzoni:2006ky,Becher:2006nr,Becher:2007ty,Bonvini:2014qga} in the way the threshold limits are defined.              
In the latter,
resummation is done in Mellin-Fourier space spanned by $(N,M)$, which corresponds to the 
scaling variable $z$ and the partonic rapidity $y_p$.  By taking the limit $N\to \infty$ and keeping $M$ fixed,  
the resummed result turns out to be identical to the inclusive one. 

%


\textit {Theoretical framework}.---The rapidity distribution of the state $F$ can be written as
\begin{eqnarray}\label{sighad}
{d \sigma^I\over dy } &=&
\sigma^I_{\rm B}(x_1^0,x_2^0,q^2,\mu_R^2) 
\sum_{ab=q,\overline q,g}
\int_{x_1^0}^1 {dz_1 \over z_1}\int_{x_2^0}^1 {dz_2 \over z_2}~ 
\nonumber \\
&&
\!\!\times
{\cal H}^I_{ab}\left({x_1^0 \over z_1},{x_2^0\over z_2},
\mu_F^2\right)
\Delta^I_{d,ab} (z_1,z_2,q^2,\mu_F^2,\mu_R^2).
\end{eqnarray}
In the above, $\mu_R$ is the ultraviolet renormalization scale,
the hadron level rapidity  
$y={1 \over 2} \ln(p_2.q/p_1.q)={1 \over 2 } \ln\left({x_1^0/x_2^0}\right)$ ~~and $\tau=q^2/S
=x_1^0 x_2^0$,
$q$ being the momentum of the final state $F$, $S=(p_1+p_2)^2$, 
where $p_i$ are the momenta of
incoming hadrons $P_i~(i=1,2)$.
For the DY process ($I=q$), 
the state $F$ is a pair of leptons with invariant mass $q^2$,
$\sigma^I=d\sigma^{q}(\tau,q^2,y)/dq^2$ whereas for
the Higgs boson production through gluon (bottom-antibottom)
fusion [$I=g(b)$], $\sigma^I=\sigma^{g(b)}(\tau,q^2,y)$.   
The function ${\cal H}^I_{ab}$ in Eq.({\ref{sighad}}) is given by
\begin{align}
{\cal H}^I_{ab}(x_1,x_2,\mu_F^2)&=
f^{P_1}_a(x_1,\mu_F^2)~ f^{P_2}_{b}(x_2,\mu_F^2)\,,
\end{align}
where $f^{P_1}_a(x_1,\mu_F^2)$ and $f^{P_2}_b(x_2,\mu_F^2)$ are the PDFs  
with momentum fractions $x_i~(i=1,2)$,
renormalized at the factorization scale $\mu_F$. 
The partonic coefficient functions, $\Delta^I_{d,ab}$,
depend on the parton-level scaling variables $z_i=\frac{x_i^0}{x_i},i=1,2$. 

Using factorization properties of the cross sections and renormalization group invariance,
in Ref.~\cite{Ravindran:2006bu},  the threshold-enhanced contribution to the $\Delta_{d,ab}^I$ 
denoted by $\Delta^{\rm SV}_{d,I}$
was shown to exponentiate as  
\begin{align}\label{delta}
\Delta^{\rm SV}_{d,I} ={\cal C} \exp
\Big({\Psi^I_d(q^2,\mu_R^2,\mu_F^2,\zo,\zt,\epsilon)}\Big)\, \Big|_{\epsilon = 0} \,,
\end{align}
where the exponent $\Psi^I_d $ is both ultraviolet and infrared finite to all orders in perturbation theory.
It contains finite distributions
computed in $4+\epsilon$ space-time dimensions expressed in terms of two shifted scaling variables 
$\zo = 1- z_1$ and $\zt = 1-z_2$, 
\begin{align} \label{psi}
\Psi^I_d &=
\Big(
\ln \Big(Z^I(\hat a_s,\mu_R^2,\mu^2,\epsilon)\Big)^2
\nonumber\\
& 
+ \ln \big|\hat F^I(\hat a_s,Q^2,\mu^2,\epsilon)\big|^2
\Big)
\delta(\zo) \delta(\zt)
\nonumber\\
& 
- {\cal C} \Big( \ln \Gamma_{II}(\hat a_s,\mu^2,\mu_F^2,\zo,\epsilon)~ \delta(\zt) + (\zo \leftrightarrow \zt)    \Big) 
\nonumber\\
& 
+ 2~ \Phi^{I}_d(\hat a_s,q^2,\mu^2,\zo,\zt,\epsilon)
,
\end{align}
where $Q^2=-q^2$ and
the scale $\mu$ is introduced to define the dimensionless strong coupling 
constant $\hat a_s=\hat g_s^2/16 \pi^2$ in dimensional regularization,
which is related to the renormalized one, $a_s$ through the 
renormalization constant $Z (a_s(\mu_R^2))$. 
The definition of double Mellin convolution 
${\cal C}$ is given in Ref.~\cite{Ravindran:2006bu}.
The overall operator renormalization constant
$Z^I$  renormalizes the bare form factor $\hat F^I$; the corresponding anomalous dimension
is denoted by $\gamma_I$ and the diagonal mass factorization kernels $\Gamma_{II}$ remove the collinear
singularities.  We have factored out $\hat F^I$ and $\Gamma_{II}$ 
in $\Delta^{\rm SV}_{d,I}$ in such a way that the remaining soft distribution function, $\Phi^{~I}_d$, 
contains only soft gluon contributions. 
Both $\hat F^I$ and $\Phi^{I}_d$ 
satisfy Sudakov-type differential equations (suppressing the arguments 
$\hat a_s,\mu^2,\zo,\zt$ for brevity), 
\begin{align}
&\chi^2 {d \over d\chi^2}\Pi^{~I}_d =
{1 \over 2 }
\Bigg[{K}^{I}_{d,\Pi}\left(\mu_R^2,\epsilon\right)  
+ {G}^{I}_{d,\Pi}  
\left(\chi^2,\mu_R^2,\epsilon\right)\Bigg]\,,
\end{align}
where $\chi^2=Q^2$ for $\Pi^I_d=\ln \hat F^I$ and $\chi^2=q^2$ for $\Pi^I_d=\Phi^I_d$.
The constants ${K}^{I}_{d,\Pi}  (\mu_R^2,\epsilon )$  
contain singular terms in $\epsilon$ and the $G^{I}_{d,\Pi}  (\chi^2,\mu_R^2,\epsilon )$
are finite in $\epsilon$.
It is straightforward to solve the above
differential equations in powers of $a_s$ and they can be
found in Refs.~\cite{Ravindran:2005vv,Ravindran:2006cg,Ravindran:2006bu,Ahmed:2014uya}. 
Substituting these solutions in Eq.(\ref{psi}) and setting $\mu_R^2=\mu_F^2$, we find 
\begin{eqnarray}
\Psi^I_d &&= ~\delta(\overline z_2)~ \Bigg({ 1 \over \overline z_1} \Bigg\{ \int_{\mu_F^2}^{q^2 ~\overline z_1}
{d \lambda^2 \over \lambda^2}~ A_I\left(a_s(\lambda^2)\right) 
\nonumber\\&&
+ D^I_d\left(a_s(q^2~\zo)\right) \Bigg\}
\Bigg)_+ + {1 \over 2} \Bigg( {1 \over \overline z_1 \overline z_2 }
\Bigg\{A^I(a_s(z_{12})) 
\nonumber \\
&& + {d D^{I}_d(a_s(z_{12}))\over d\ln z_{12}} \Bigg\}\Bigg)_+
+ {1 \over 2}
\delta(\overline z_1) \delta(\overline z_2)
\ln \Big(g^I_{d,0}(a_s(\mu_F^2))\Big)
\nonumber\\&&
+ \overline z_1 \leftrightarrow \overline z_2,
\end{eqnarray}
where the subscript $+$ indicates the standard plus distribution, $A^I$ are cusp anomalous dimensions and the constants $z_{12}=q^2 \overline z_1 \overline z_2$.
The finite function $D^I_d$ is defined through $G^I_{d,\Phi}\left(q^2,z_i,\epsilon\right)$ in
the limit $\epsilon\to 0$ expanded in $a_s$ as 
\begin{eqnarray}
\label{eqn:DId}
D^I_d( a_s(q^2 z_i)) 
&=&\sum_{j=1}^\infty a_s^j\left(q^2 z_i \right)
G^{I,(j)}_{d,\Phi}\left(q^2,z_i,\epsilon\right)\bigg|_{\epsilon=0}
\nonumber\\
&=&\sum_{j=1}^\infty a_s^j\left(q^2 z_i \right) 
\bigg( C^I_{d,j} -f^I_j + \sum_{k=1}^{\infty} \epsilon^k \overline{{\cal G}}_{d,j}^{I,k}\bigg)\bigg|_{\epsilon=0}
\end{eqnarray}
%
%
%
The constants $C^I_{d,j}$ can be expressed in terms of lower order $\overline {\cal G}^{I,k}_{d,j}$ 
(see Eq.(32) of Ref.~\cite{Ravindran:2006bu}), and  
the soft anomalous dimensions $f_j^I$ are known up to three loops (see Refs.~\cite{Ravindran:2004mb,Moch:2005id}).
The constants $\overline {\cal G}^{I,k}_{d,j} $ and hence $D^I_d$ in Eq.(\ref{eqn:DId}) can be determined using %
\begin{align}
\int_0^1 
dx_1^0 \int_0^1 
dx_2^0 \left(x_1^0 x_2^0\right)^{N-1}
{d \sigma^I \over d y}
=\int_0^1 d\tau~ \tau^{N-1} ~\sigma^I\,,
\label{iden}
\end{align}
where the $\sigma^I$ is the inclusive cross section.  
Since we are interested in the threshold limit,
we consider the limit $N\rightarrow \infty$ on both sides and use the well-known threshold resummed
inclusive cross section, $\sigma^{I,\text{res}}$, in $N$ space to obtain the unknown constants
$\overline {\cal G}^{I,k}_{d,i}$ and hence the unknown $D_{d}^I$.   
Alternatively, we can use the $z$-space approach to determine 
these constants in terms of the corresponding ones from the inclusive cross section 
as they are independent of scaling variables $z_i$ and $z$.  
Hence, using 
the $z$-space formalism for the inclusive cross section described in Refs.~\cite{Ravindran:2005vv,Ravindran:2006cg} 
and for rapidity distribution 
in Ref.~\cite{Ravindran:2006bu}, we can express  
$\overline {\cal G}^{I,k}_{d,i}$ 
in terms $\overline {\cal G}^{I,k}_{i}$.
%
Substituting these constants in Eq.(\ref{eqn:DId}), expanding 
$D^I_d$ in powers $a_s$ as $D^I_{d} = \sum_{j=1}^\infty a_s^j D^I_{d,j}$
and comparing against $D^I$ from the inclusive cross section, we obtain
\begin{eqnarray}
D_{d,1}^I &=&  D^I_1,
\nonumber\\
D_{d,2}^I &=& D^I_2 -\zeta_2 \beta_0 A_1^I,
\nonumber\\
D_{d,3}^I &=& D^I_3 - \zeta_2(\beta_1 A_1^I +2 \beta_0 A_2^I
+2 \beta_0^2 f_1^I ) - 4 \zeta_3 \beta_0^2 A_1^I 
\end{eqnarray}
From the above equations it is clear that $D^g_{d,j} = D^q_{d,j} C_A/C_F, j=1,2,3$, i.e., maximally non-Abelian. 
Following 
Ref.~\cite{Catani:2003zt} and 
defining $\w = a_s \beta_0 \ln(\overline N_1 \overline N_2)$ where
$\overline N_i = e^{\gamma_E} N_i, i=1,2$, we find
\begin{eqnarray}
\label{eqn:expG}
\tilde \Delta_{d,I}^{\text{SV}}(\w) &=& \int_0^1 dz_1 z_1^{ N_1-1} \int_0^1 dz_2  z_2^{ N_2-1}  \Delta_{d,I}^{\text{SV}}
\nonumber \\
&=& g^I_{d,0}(a_s) \exp\Big( g^I_d (a_s,\w)\Big), 
\end{eqnarray}
where $\gamma_E=0.57721566\cdot\cdot\cdot$ is the Euler-Mascheroni constant.
The exponent $g^I_d(a_s,\w)$ takes the canonical form:
\begin{equation}
g^I_d(a_s,\w)=g^I_{d,1}(\w) \ln(\overline N_1 \overline N_2) + \sum_{i=0}^\infty a_s^i g^I_{d,i+2}(\w)\,.  
\end{equation}
Rescaling the constants by $\beta_0$ as 
$\overline g^I_{d,1} = g^I_{d,1}$, $\overline g^I_{d,i+2} = g^I_{d,i+2}/\beta_0^i$, 
$\overline A_i^I = A_i^I /\beta_0^i $, $\overline D_{d,i}^I = 
D_{d,i}^I/ \beta_0^i$ and $\overline \beta_i = \beta_i /\beta_0^{i+1}$, we find
\begin{widetext}
\begin{eqnarray}
\label{eqn:G}
\overline g^I_{d,1} &=& \Aob {1 \over \w} \Big(\w +(1-\w) \lw\Big) ,
\nonumber\\
\overline g^I_{d,2} &=& \w\left(  \Aob \btob -\Atb  \right)+ \ln(1-\w)\left( \Aob \btob + \Dobd - \Atb \right)+ {1\over2}\ln^2(1-\w) \Aob\btob   +  \lqr\ln(1-\w)\Aob + \lfr\w \Aob , 
\nonumber\\
\overline g^I_{d,3} &=&  - {\w\over2}\Athb - {\w \over 2 (1-\w)} \left( -\Athb+(2+\w)\btob\Atb+\left( (\w-2)\bttb-\w\btob^2-2\zeta_2\right)\Aob+2\Dtbd-2\btob\Dobd \right)
\nonumber\\&&
-\lw\left({\btob\over(1-\w)}\Big( \Atb-\Dobd-\Aob\btob\w\Big)-\Aob\bttb  \right)+{\ln^2(1-\w)\over2(1-\w)}\Aob\btob^2 +  \lfr \Atb\w- {1 \over 2}\lfr^2 \Aob  \w
\nonumber\\&&
 - \lqr{1\over(1-\w)} \left(\left(\Atb -\Dobd\right)\w-\Aob\btob\left(\w+\lw \right)\right) +{1 \over 2}\lqr^2  {\w \over  (1-\w)} \Aob.
\end{eqnarray}
\end{widetext}
where $\lfr = \ln(\mu_F^2/\mu_R^2), \lqr = \ln(q^2/\mu_R^2)$.
Expanding $\ln(g^I_{d,0})$ as
$\ln(g^I_{d,0}) = \sum_{i=1}^\infty a_s^i  l^{I,(i)}_{g_0}$,
we find
\begin{widetext}
\begin{eqnarray}
l^{I,(1)}_{g_0} &=& 2G_1^{I,1} + 2 \overline {\cal G}_{d,1}^{I,1} + 4 \Ao \zeta_2 - 2\lfr\Bo+2\lqr\left(\Bo-\gamma^I_0\right),
\nonumber\\
l^{I,(2)}_{g_0} &=& G_2^{I,1} + \overline {\cal G}_{d,2}^{I,1} +2\beta_0\left(G_1^{I,2}+\overline {\cal G}_{d,1}^{I,2} \right)+2\zeta_2\left(2\At+\beta_0\left( 3\Bo+2f^I_1-3\gamma^I_0\right) \right)+{2\over3}\Ao\beta_0\zeta_3 -2 \lfr\Bt  + \lfr^2 \Bo \beta_0 
\nonumber\\&&
 +  \lqr \Big(2 \Bt - 2 \gamma^I_1 - \beta_0 \Big(2 G_1^{I,1} + 2 \overline {\cal G}_{d,1}^{I,1} + 
4 \Ao \zeta_2\Big)\Big)+ \lqr^2\beta_0 \Big(-\Bo + \gamma^I_0\Big).
\end{eqnarray}
\end{widetext}
The expression for $\overline g^I_{d,4}$ and $l^{I,(3)}_{g_0}$ can be found in Ref.~\cite{Banerjee:2017cfc}, the online version of this paper.
In the above equation, $G^{I,k}_j$s are obtained from the $\epsilon$-dependent part of $G^I_{d,F}$, 
and $B^I_j$ are the coefficients of $\delta(1-z)$ in 
$\Gamma_{II}$.    
The all-order resummed result given in Eq.~(\ref{eqn:expG}) 
is the main result of this paper.
%
Exponentiation of the functions $g^I_{d,i}$ resums the terms $a_s \beta_0 \ln(\overline N_1 \overline N_2)$ 
systematically 
to all orders in perturbation theory
analogous to the inclusive one (see Ref.~\cite{Catani:2003zt}).
The resummed result 
can be used to study the rapidity distribution of any colorless particle $F$ produced in hadron-hadron collision. 
In this paper, we restrict ourselves to the production of a scalar Higgs boson at the LHC and present the numerical impact
of the resummed result over the fixed-order result known to NNLO level \cite{Anastasiou:2005qj}.
This is obtained using 
\begin{align}
\label{eqn:Reseq}
{d \sigma^{g,\text{res}} \over dy } &=  
{d \sigma^{g,\text{f.o}} \over dy } +
\, {\sigma^{g}_B } \int_{c_{1} - i\infty}^{c_1 + i\infty} \frac{d N_{1}}{2\pi i}
 \int_{c_{2} - i\infty}^{c_2 + i\infty} \frac{d N_{2}}{2\pi i} 
\nonumber\\&
\times
e^{y(N_{2}'-N_{1}')}
\left(\sqrt{\tau}\right)^{-2-N_{1}'-N_{2}'} 
\tilde f_{g}(N_{1}') 
 \tilde f_{g}(N_{2}') 
\nonumber\\&
\times
\Bigg[ \tilde \Delta_{d,g}^{\text{SV},N_1',N_2'} -  \left(\tilde \Delta_{d,g}^{\text{SV},N_1',N_2'}\right)_{\text{trunc}} \Bigg]  
\end{align}
where $N_i'=N_i+1, i=1,2$.  In the above equation, 
the superscript "f.o" refers to the fixed-order 
result in $a_s$ and    
"res" refers to the resummed result. 
The subscript "trunc" refers to the result obtained
from Eq.~(\ref{eqn:expG}) by truncating at desired accuracy in $a_s$. 
The constants $g_{d,0}^g$ and $g_{d,i}^g$ 
that appear in $\tilde \Delta_{d,g}^{\text{SV}}$
are functions of cusp ($A^g_i$), collinear ($B^g_i$), 
soft ($f^g_i$), UV ($\gamma^g_i$) anomalous dimensions and 
universal soft terms ${\overline{\mathcal G}}^{g,i}_{d,j}$, and process-dependent constants $G^{g,i}_j$ of virtual corrections, and they 
are known to next-to-next-to-leading-logarithmic (NNLL) accuracy.
We performed double Mellin inversions
to obtain the final result in terms of $\tau$ and $y$ and used minimal prescription advocated in
Ref.~\cite{Catani:1996yz}.
For the resummed result to N$^m$LO+ N$^n$LL we need f.o to N$^m$LO accuracy and 
$\tilde \Delta^{\text{SV}}_{d,g}$ to N$^n$LL accuracy.  For the latter, 
we need $g^g_{d,0}$ 
up to order $a_s^n$, and for the exponent, we need all the terms up to $g^{g}_{d,n+1}$.

\begin{figure}[t!]
\centerline{
\includegraphics[width=9cm]{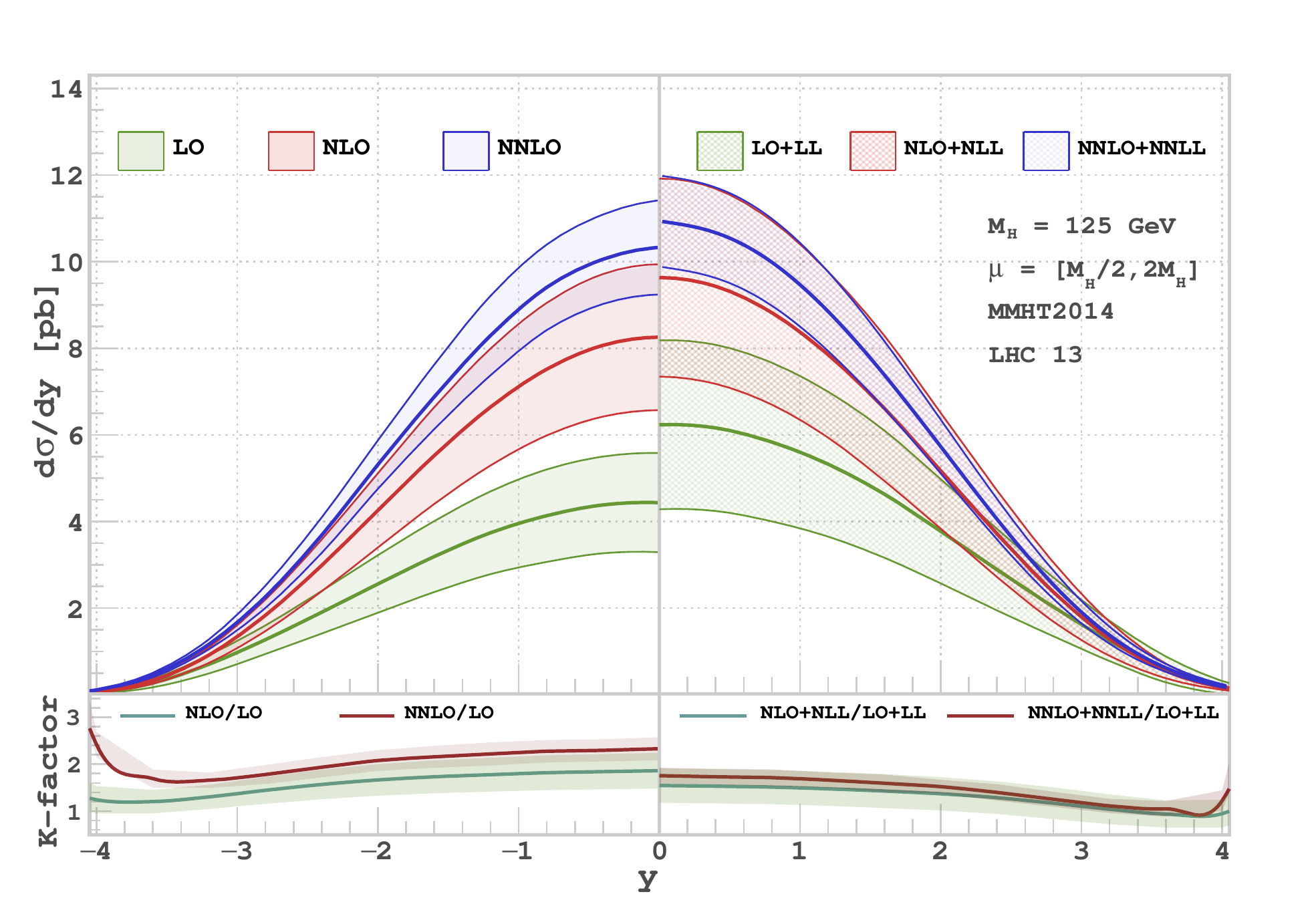}
}
\caption{Higgs rapidity distributions for fixed-order (left panel) and resummed contributions (right panel)
are presented with corresponding $K$ factors on lower panels around the central scale $\mu_R=\mu_F=M_H$.}
\label{fig1}
\end{figure}
 \textit {Phenomenology}.--- In the following we study the numerical impact of resummed contributions up to NNLL accuracy for the rapidity distribution of a scalar Higgs boson of mass $M_H=125$ GeV at the LHC with $\sqrt{S} = 13$ TeV. We have set the number of flavors $n_f=5$ and the top mass at 
173 GeV and use MMHT 2014 \cite{Harland-Lang:2014zoa} PDFs along with the corresponding values of $a_{s}$ for LO, NLO, and NNLO through the LHAPDF \cite{Buckley:2014ana} interface, unless otherwise stated. We use the publicly available code FEHIP~\cite{Anastasiou:2005qj} to obtain $d\sigma^{g,\text{f.o}}/dy$ up to NNLO level.  We have developed an in-house Fortran code to perform double Mellin inversion for the resummed contributions computed in this paper. In Fig.\ref{fig1}, using Eq.(\ref{eqn:Reseq}), we present the production cross section for the scalar Higgs boson as a function of its rapidity $y$ up to NNLO in the left panel and to NNLO+NNLL in the right panel along with the respective K factors.  
The K factor at a given order, say, at N$^n$LO (N$^n$LO + N$^n$LL), is defined by the cross section at that order normalized by the same at LO (LO+LL) at $\mu_R=\mu_F=M_H$.  The symmetric band at each order is generated by varying 
$\mu_R$ and $\mu_{F}$ between $[M_{H}/2, 2 M_{H}]$ around the central scale $\mu_R=\mu_F=M_H$ with the 
constraint $1/2\leq \mu_R/ \mu_F\leq 2$, 
adding and subtracting the highest possible errors from all the scale combinations to the central scale.
We find that the magnitude and sign of the resummed contribution do vary depending on the order in $a_s$ as well the exact values of $y$ and the scales $\mu_R,\mu_F$.

\begin{widetext}
\begin{center}
\begin{table}
 \renewcommand{\arraystretch}{1.5}
\begin{tabular}{ |P{1.3cm}|P{2cm}|P{2cm}|P{2cm}|P{2cm}|P{2cm}|P{2.1cm}|P{2.3cm}|}
 \hline
  y &LO&LO+LL&NLO&NLO+NLL&NNLO  & NNLO+NNLL  & NNLO+NNNLL\\
 \hline
  0.0 &$4.435\pm 1.145$&$6.231\pm 1.950$&$8.255\pm 1.684$& $9.632\pm 2.286$ &$10.329\pm 1.088$   &$10.938\pm 1.050$&$10.517\pm 0.820$\\
  \hline
  0.8 & $4.134\pm 1.067$  &$5.833\pm 1.831$ &$7.517\pm 1.530$&$8.820\pm 2.124$&$9.407\pm 0.988$ &$9.992\pm 1.025$& $9.641\pm 0.718$\\
  \hline
  1.6 &$3.189\pm 0.819$&$4.630\pm 1.468$&$5.522\pm 1.117$&$6.611\pm 1.676$&$6.877\pm 0.744$ &$7.380\pm 0.849$ &$ 7.045\pm0.563$\\
  \hline
  2.4& $1.904\pm 0.492$&$2.887\pm 0.942$ &$2.985\pm 0.597$&$3.715\pm .998$&$3.683\pm 0.410$&$4.040\pm 0.501$&$3.821\pm 0.305$\\
\hline
\end{tabular}
\caption{Fixed-order and resummed results for Higgs rapidity distribution with corresponding absolute error for different benchmark values of $y$.} \label{table1}
\end{table}
\end{center}
\end{widetext}
  In particular, at the central scale $\mu=\mu_R=\mu_F=M_H$, the percentage correction from the leading-logarithmic (LL) contribution 
goes from $40 \%$ to $50\%$ whereas for next-to-leading-logarithmic (NLL), we find that it varies from $17\%$ to $24\%$ and for
NNLL it varies from $6\%$ to $10\%$ in the region $0\le y\le 2.4$, which is evident from Table~\ref{table1}.      
Interestingly, at $\mu = M_{H}/2$, 
we find that the cross section at NNLO+NNLL is very close to NNLO for a wider
range of $y$ indicating that $\mu=M_H/2$ is a good choice for the fixed-order predictions.
A similar conclusion was arrived at in Ref.~\cite{Anastasiou:2003yy} for the inclusive production of the Higgs boson. From the upper-left panel of Fig.\ref{fig1}, we also observe that LO and NLO predictions do not overlap around the central rapidity region. However, at NNLO, partial overlap indicates that the inclusion of higher-order corrections has increased the convergence of perturbation series. The upper-right panel shows the effect of resummation over the fixed-order result. We observe that LO+LL has overlap 
with NLO+NLL for all values of rapidity. In addition, the distribution at NNLO+NNLL  falls completely within  NLO+NLL band. In fact, NNLO+NNLL  increases approximately by 13\% with respect to NLO+NLL; the corresponding number for NNLO over NLO is approximately 25\%. This implies that the perturbative convergence at the resummed level is better compared to the fixed-order result. We have also chosen $M_H/2$ as the central scale and found out that the choice of central scale has a minimum effect on the resummed result at NNLO+NNLL; i.e., the resum result at this order stabilizes irrespective of the above-mentioned choices, whereas at fixed order, this does not happen.  Based on the above observations, 
we can predict that the N$^3$LO will be very close to NNLO+NNLL and the N$^3$LO + N$^3$LL result will lie within the NNLO+NNLL uncertainty band.  In the Table~\ref{table1}, the impact of N$^3$LL on the NNLO result is also presented.
   
\begin{figure}[t!]
\centerline{
\includegraphics[width=9cm]{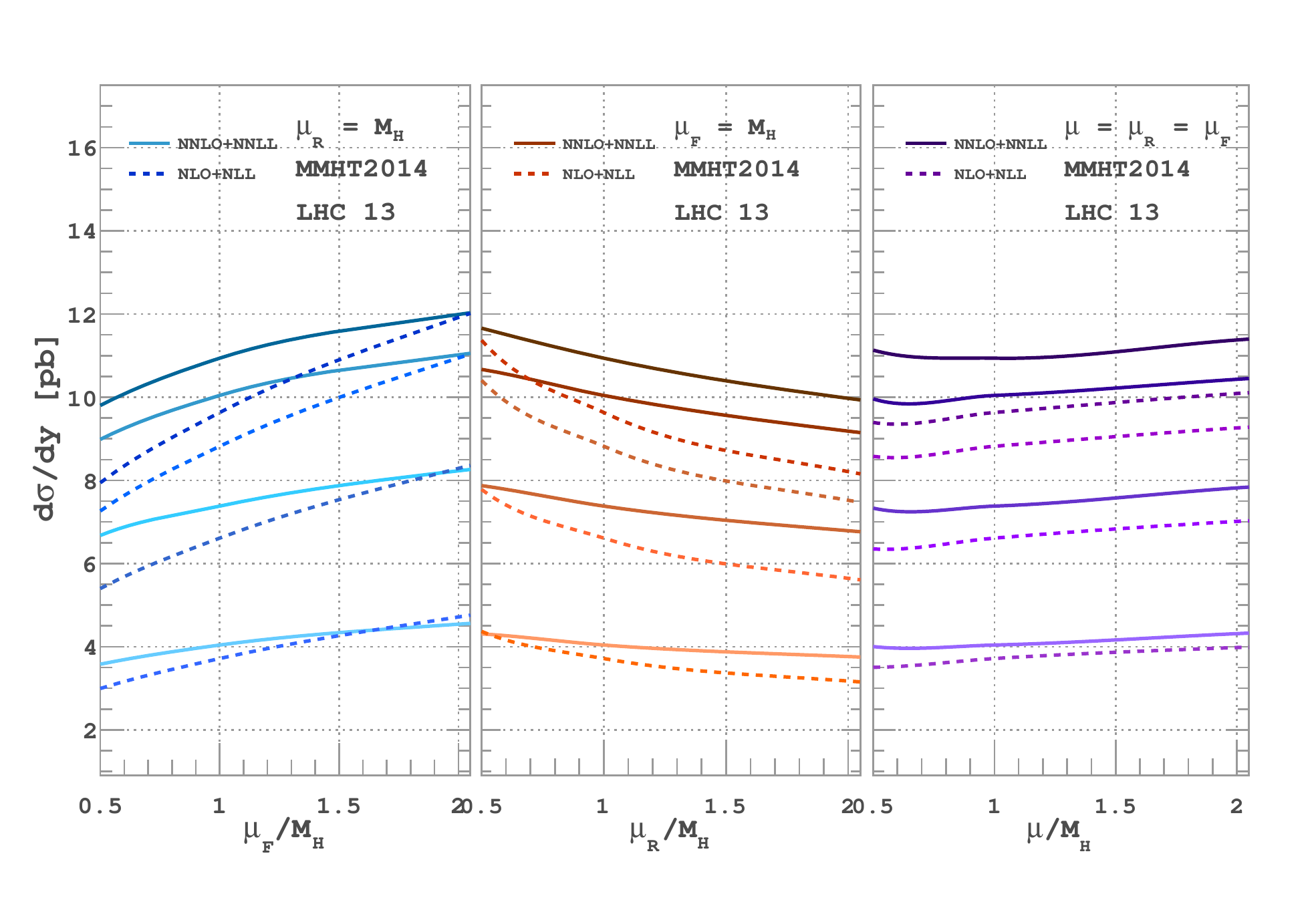}
}
\caption{$\mu_F$, $\mu_R$ scale variations for the NLO+NLL (dashed) and NNLO+NNLL (solid) cases for different benchmark $y$ values (starting from the top, $y=0, 0.8, 1.6, 2.4$).}
\label{fig2}
\end{figure}

To understand the impact of unphysical scales $\mu_R$ and $\mu_F$ on our resummed results, we first varied 
one while fixing the other to $M_H$ and then varied both simultaneously for various values of rapidity $y$, the results
are presented in Fig.~\ref{fig2}.  As expected, the running coupling constant decreased the cross section as 
we increased $\mu_R$, while the opposite behaviour was observed for $\mu_F$ both in fixed-order and in resummed results. 
Varying these two scales simultaneously led to a cancellation of the two different behaviors, and the amount 
of cancellation depended on order of perturbation $n$ and value of $y$.   
Finally, to study the impact of choice of PDFs, in Table~\ref{table2}, we have presented 
the results at NNLO+NNLL using the central PDF of each PDF group.

\begin{table}
\renewcommand{\arraystretch}{1.5}
\begin{tabular}{ |P{0.5cm}|P{1.3cm}|P{1.3cm}|P{1.3cm}|P{1.3cm}|P{1.5cm}| }
 \hline
  y &MMHT  & ABMP & CT10 &NNPDF & PDF4LHC  \\
 \hline
  0.0 &  $10.938$   &$10.654$ &$10.709$ & $11.302$ &$10.850$  \\
  \hline
  0.8 &  $9.992$  & $9.713$  & $9.820$& $10.378$ &$9.977$ \\
  \hline
  1.6 &$7.380$ &$7.043$ &$7.362$ &$7.758$ &$7.456$  \\
  \hline
  2.4&$4.040$ &$3.727$ &$4.105$ & $4.111$ &$4.075$  \\
\hline
\end{tabular}
\caption{Using different PDFs, NNLO+NNLL contributions to rapidity distribution for  $y=0, 0.8, 1.6, 2.4$.}
\label{table2}
\end{table}

\textit{Conclusion}.---In this paper, we have developed a formalism to resum threshold logarithms in double Mellin space for the rapidity distribution of a colorless final state $F$ produced in the hadron collider. We have derived for the first time compact and most general expressions for resummed exponents $g^{I}_{d}$ up to NNLO+NNLL accuracy. We find that the resummed result not only changes the fixed-order predictions but also remarkably improves the perturbative convergence.  We observe that the resummed result at NNLO+NNLL stabilizes over fixed order irrespective of the choices of the central scale between $[M_H/2, 2M_H]$. We have also studied the impact of PDFs on the predictions. The present study can easily be extended to Drell-Yan \cite{Pulak:dy}, pseudoscalar, and $W^\pm$ and $Z$ productions as well as the production of the Higgs boson in bottom-antibottom annihilation at hadron colliders.  

\textit {Acknowledgements}.---
We are thankful for useful help from 
A. Vogt,
F. Petriello, 
T. Hahn, 
M. Bonvini, 
L. Rottoli,
P. Mangalapandi, 
T. Ahmed,
N. Rana,
and 
A. Karan. V. Ravindran would like to thank T. Gehrmann for fruitful discussions 
and also thanks University of Zurich, where part of the work was carried out, for hospitality.

\bibliography{Higgsres}

\bibliographystyle{apsrev4-1}
\end{document}